\documentstyle[aps,epsfig,eqsecnum,prb]{revtex}

\begin{document}
\title{Suppression of current in transport through parallel double quantum dots}
\draft
\author{Tae-Suk Kim$^{a,b}$ and S. Hershfield$^{b}$}
\address{$^a$ Institute of Physics and Applied Physics, Yonsei University, 
  Seoul 120-749, Korea \\
 $^b$ Department of Physics, University of Florida, Gainesville FL 32611-8440}
\date{\today}
\maketitle
\begin{abstract}
 We report our study of the I-V curves in the transport through the quantum dot 
when an additional quantum dot lying in the Kondo regime is side-connected to it.
Due to the Kondo scattering off the effective spin on a side-connected quantum dot 
the conductance is suppressed at low temperatures and at low source-drain bias
voltages. This zero-bias anomaly is understood as enhanced Kondo scattering 
with decreasing temperature.

\end{abstract}
\pacs{72.10.Fk, 72.15.Qm}

\section{Introduction}
 Advances in nanotechnology have made it possible to fabricate quantum dots
or artificial atoms and to study their transport properties. 
Due to the confinement and the Coulomb interaction in the quantum dot, 
charge and spin can be quantized. Quantization of electron
charge leads to the almost periodic variation of conductance as a function of
the number of electrons $N$ or the gate voltage. A conductance peak is observed 
whenever the charge fluctuation is allowed or the $N$- and $(N+1)$-states 
become energetically degenerate. Otherwise electrons in the electrodes cannot 
hop into a quantum dot due to the strong Coulomb repulsion.  
This Coulomb blockade leads to 
vanishing conductance between two neighboring conductance peaks. 
Novel Kondo effects can enhance the conductance in the Coulomb blockade region 
when the number of electrons at the quantum dot is odd and at least one 
electron spin is unpaired. The unpaired $S=1/2$ in the quantum dot 
can be screened by the electrons in the external electrodes connected 
to the quantum dot. This theoretical prediction\cite{thekondo} of 
the novel nonequilibrium Kondo effects in quantum dot systems was confirmed 
experimentally recently\cite{expkondo1,expkondo2,expkondo3}. 
In the conventional bulk Kondo systems where magnetic ions 
are doped into normal metals the study of the Kondo effects was confined to 
the equilibrium state. In quantum dot systems the nonequilibrium situation
is easily controlled and model parameters can be varied by adjusting 
the gate voltages.

 The study of the Kondo effect in the quantum dot systems has been diversified
to see the different aspects.
 The quantum dots are inserted in the Aharonov-Bohm ring
to study the effect of many-body interaction on the persistent current\cite{abring}. 
When electrons are scattered off the magnetic impurity, they experience a phase
shift $\pi/2$ at the Fermi energy due to the Kondo effect. The direct measurement of
this phase shift is not possible in the bulk systems. Using a quantum dot 
inserted in the Aharonov-Bohm ring the phase shift due to the Kondo effect 
was studied theoretically\cite{thephase} 
and measured experimentally\cite{expphase}. 
Serial double quantum dots were also studied in the Kondo regime\cite{sddot}. 
Enhanced conductance was observed\cite{expeven} 
in an otherwise Coulomb blockade region 
at the spin singlet-to-triplet transition point when the 
number of electrons on a quantum dot is even. The ground state can be 
either spin-singlet or spin-triplet depending on the geometry of the confinement 
potential. Perpendicular magnetic fields can induce a transition of the ground state
between spin-singlet and spin-triplet. This novel Kondo effects have been interpreted
as coming from the enhanced Kondo temperature at the transition point\cite{theeven}.

\begin{figure}
\protect\centerline{\epsfxsize=4.0in \epsfbox{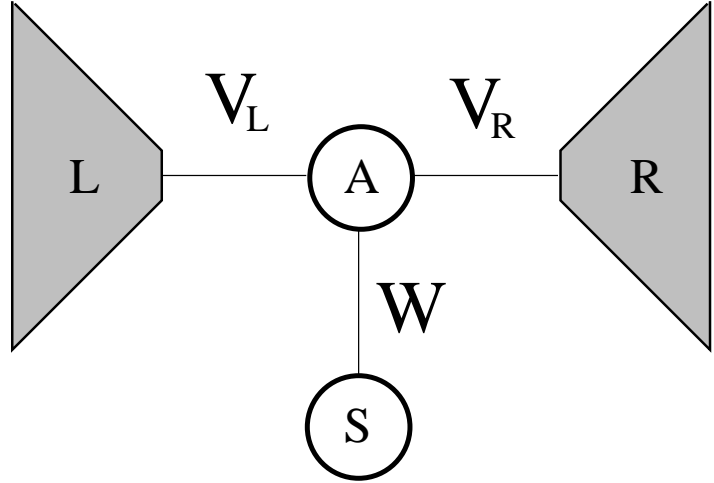}}
\vskip 0.5cm
\protect\caption[Fig.1.]
{{\bf Schematic display of parallel double quantum dots.} The {\it active} 
quantum dot $A$ is connected to two left and right electrodes and 
the other dot $S$ is {\it side-connected} to the dot $A$. 
A side-connected dot $S$ provides an additional current path 
or acts as scattering center of electrons
passing through the dot $A$.
}
\protect\label{pddot}
\end{figure}

 In this paper we study theoretically the transport properties of parallel 
double quantum dots as depicted in Fig.\ref{pddot}. 
One {\it active} quantun dot is connected to the source and drain 
electrodes and the other quantum dot is {\it side-connected}
to the active quantum dot. 
Since the energy levels in two quantum dots can be controlled separately using 
the gate voltage, different transport regimes can be probed. 
In this work we present our study of the 
I-V characteristics of this system when the active quantum dot lies in the 
conductance peak region while the side-connected quantum dot in the Coulomb 
blockade region. In this case charge fluctuations are suppressed and 
an effective spin $S=1/2$ arises in a side-connected dot.
Electrons passing through the active dot 
experience the Kondo scattering off an effective spin in a side-connected dot.
We find that the linear response conductance is suppressed at low temperature 
due to the enhanced Kondo scattering off spin $S=1/2$. 
The spectral function at a side-connected dot develops a Kondo resonance 
peak as the temperature is lowered. On the other hand the spectral function 
at the active dot becomes depleted near $\omega=0$ with decreasing temperature.  
We obtained these results using the nonequilibrium Green's function 
method combining with the non-crossing approximation (NCA).

 This paper is organized as follows. In section II we introduce our model 
Hamiltonian and present the formulation of current passing through an active
dot. The relevant Green's functions in nonequilibrium are formulated in section III
using the non-crossing approximation (NCA) which is modified for our system. 
Numerical results, solving the NCA equations self-consistently, are presented 
in section IV and conclusion is included in section V.

\section{Parallel Double Quantum Dots}
\subsection{Model Hamiltonian}
 Consider the parallel double quantum dots shown in Fig.~\ref{pddot}. 
In this experimental geometry, the transfer of electrons from one electrode 
to the other is realized by hopping on and off an {\it active} quantum dot 
(the site $A$). The other dot (the site $S$) is {\it side-connected} 
to the active dot by hopping. 
Assuming that the energy level spacing is large enough 
compared to the level-broadening due to the tunneling into two electrodes,
the model Hamiltonian can be written as
\begin{mathletters}
\begin{eqnarray}
H &=& H_{\rm el} + H_{\rm qdot} + H_1, \\
H_{\rm el}
 &=& \sum_{p=L,R} \sum_{\vec{k}\alpha} \epsilon_{p\vec{k}} ~
  c_{p\vec{k}\alpha}^{\dag} c_{p\vec{k}\alpha}^{\phantom{\dag}}, \\
H_{\rm qdot} 
 &=& \sum_{i=A,S}\sum_{\alpha} E_{i} ~d_{i\alpha}^{\dag} d_{i\alpha}^{\phantom{\dag}} 
   + \sum_{i=A,S} U_i n_{i\uparrow} n_{i\downarrow}, \\
H_1 &=& \sum_{p=L,R} \sum_{\vec{k}\alpha} \left[ 
    V_{p}(\vec{k}) ~c_{p\vec{k}\alpha}^{\dag} d_{A\alpha}^{\phantom{\dag}}
    + V_{p}^{*}(\vec{k}) ~d_{A\alpha}^{\dag} c_{p\vec{k}\alpha}^{\phantom{\dag}}
   \right] 
 + \sum_{\alpha} \left[ W d_{A\alpha}^{\dag} d_{S\alpha}^{\phantom{\dag}}
      + W^* d_{S\alpha}^{\dag} d_{A\alpha}^{\phantom{\dag}} \right]. 
\end{eqnarray}
\end{mathletters}
The electron creation operator of spin $\alpha$ in the electrode $p=L,R$
is $c_{p\vec{k}\alpha}^{\dag}$.
The index $p=L(R)$ denotes the left(right) external electrode. Both electrodes 
are assumed to be described by the Lorentzian density of states (DOS) with the 
energy dispersion $\epsilon_{p\vec{k}} \equiv \epsilon_{\vec{k}} + \mu_p$. 
$\mu_{L,R} = \pm {1\over 2} eV$ is the chemical potential shift due to the bias
voltage applied to the electrodes. 
$d_{i\alpha}^{\dag}$ is the electron creation operator in a quantum dot $i=A,S$
with $A(S)$ labeling a active (side-connected) dot, respectively. 
$U_i$ is the on-site electron-electron Coulomb interaction in a dot $i$. 

When $U_A=U_S=0$, two dots become resonant levels and the solution
can be found analytically (see Appendix A). The model becomes nontrivial
when one of on-site Coulomb interactions or both are nonzero. 
In this paper we are going to study the case of $U_A=0$ and $U_S\neq 0$ in detail.  
When $U_S$ is the largest of all the model parameters and $-E_S \gg |W|$,
charge fluctuations in a side-connected dot can be neglected and 
the side-connected dot can be treated as an effective spin $S=1/2$. 
In this case removing the charge degrees of freedom in a side-connected dot 
(Schrieffer-Wolf transformation\cite{swolf}) 
the interaction between two dots can be derived.
\begin{eqnarray}
H_{AS} &=& J_{AS} \vec{S}_S \cdot \sum_{\alpha\beta} 
    d_{A\alpha}^{\dag} {1\over 2} \vec{\sigma}_{\alpha\beta} 
    d_{A\beta}^{\phantom{*}}, ~~~
  J_{AS} ~=~ {2|W|^2 \over -E_S} + {2|W|^2 \over U_S + E_S}. 
\end{eqnarray}
Here $\vec{S}_S = \sum_{\alpha\beta} |S;\alpha> {1\over 2} 
\vec{\sigma}_{\alpha\beta} <S; \beta|$ represents the spin degrees of freedom in a 
side-connected dot in the absence of charge fluctuations. 
Electrons, flowing from the left electrode to the right, 
at the active dot will experience the Kondo scattering off an effective spin 
of a side-connected dot. Appendix B contains the details of the $U_A \neq 0 $ and 
$U_S \neq 0$ case.

\subsection{Formulation of Current}
 The current operator can be defined as a variation of the number operators of 
electrons per unit time leaving one electrode ($p=L,R$). 
\begin{eqnarray}
\hat{I}_p 
 &\equiv& -e (-\dot{N_p}) ~=~ {e \over i\hbar} [ N_p, H] 
 ~=~ {e\over i\hbar} \sum_{\vec{k}\alpha} \left[ V_p(\vec{k}) ~
    c_{p\vec{k}\alpha}^{\dag}d_{A\alpha}^{\phantom{\dag}} - H.c.\right].
\end{eqnarray}
The measured current is given by the thermal average of the above current operator
and can be written in terms of the lesser Green's functions out of equilibrium
\cite{neqlandauer}.
\begin{eqnarray}
I_p (t) &=& -ie ~\sum_{\vec{k}\alpha} 
 \left[ V_p (\vec{k}) ~ G_{Acp}^{<} (\vec{k}; t,t)
   - V_p^{*} (\vec{k}) ~ G_{cpA}^{<} (\vec{k}; t,t)
 \right].
\end{eqnarray}
The current can be expressed in terms of Green's functions for the 
conduction electrons and quantum dots using the Dyson equations for 
the mixed Green's functions
\begin{mathletters}
\begin{eqnarray}
G_{cpA} (\vec{k}; t,t')
 &\equiv& {1\over i\hbar} ~\left< T c_{p\vec{k}\alpha}^{\phantom{\dag}}(t)
   d_{A\alpha}^{\dag}(t') \right> \nonumber\\
 &=& V_p(\vec{k})~\int_C dt_1~ G_{cp} (\vec{k}; t, t_1) ~ D_{A} (t_1, t'),\\
G_{Acp} (\vec{k}; t,t')
 &\equiv& {1\over i\hbar} ~\left< T d_{A\alpha}^{\phantom{\dag}} (t)
   c_{p\vec{k}\alpha}^{\dag} (t')\right> \nonumber\\
 &=& V_p^{*}(\vec{k})~\int_C dt_1~ D_{A} (t, t_1) ~ 
    G_{cp} (\vec{k}; t_1, t'), \\
i\hbar D_i (t,t') 
 &=& < T d_{i\alpha}^{\phantom{\dag}} (t) d_{i\alpha}^{\dag} (t') >, ~~
    i ~=~ A, S. 
\end{eqnarray}
\end{mathletters}
Using analytic continuation to the real-time axis\cite{langreth}, 
we can find the lesser part of $G_{Acp} (\vec{k}; t,t')$
\begin{eqnarray}
G_{Acp}^{<} (\vec{k}; t,t')
 &=& V_p^{*}(\vec{k})~\int dt_1~ \left[ 
    D_{A}^{<} (t, t_1) ~G_{cp}^{a} (\vec{k}; t_1, t')
   + D_{A}^{r} (t, t_1) ~G_{cp}^{<} (\vec{k}; t_1, t') \right]. 
\end{eqnarray} 
The current can be expressed as 
\begin{eqnarray}
\label{currenteqn}
I_p &=& -ie ~{1\over V} \sum_{\vec{k}\alpha} 
  \left| V_p (\vec{k}) \right|^2 ~\int {d\omega \over 2\pi} ~ \left\{   
   D_{A}^{<} (\omega) [ \Sigma_{cp}^{a} (\omega) - \Sigma_{cp}^{r} (\omega) ]
    + \Sigma_{cp}^{<} (\omega) [ D_{A}^{r} (\omega) - D_{A}^{a} (\omega) ] 
   \right\}  \nonumber\\
 &=& -ie \sum_{\alpha} \int {d\omega \over 2\pi} ~
   | D_{A}^{r} (\omega)|^2 \left\{ 
   \Sigma_{A}^{<} (\omega) [ \Sigma_{cp}^{a} (\omega) - \Sigma_{cp}^{r} (\omega) ]
    + \Sigma_{cp}^{<} (\omega) [ \Sigma_{A}^{r} (\omega) - \Sigma_{A}^{a} (\omega) ] 
   \right\}.  
\end{eqnarray}
Here $\Sigma_{A}$ is the self-energy for the active dot $A$ and 
the following equations are used
\begin{mathletters}
\begin{eqnarray}
\label{csigma}
\Sigma_{cp} (t,t') 
 &=& {1\over V} \sum_{\vec{k}} \left| V_p (\vec{k}) \right|^2 G_{cp} (\vec{k};t,t'), \\
\Sigma_{cp}^{r,a} (\omega) 
 &=& \int {d\epsilon \over \pi} { \Gamma_p(\epsilon) 
    \over \hbar\omega - \epsilon \pm i\delta }, \\
\Sigma_{cp}^{<,>} (\omega) 
 &=& 2 \Gamma_p(\omega) f_p (\pm\omega), \\
\Gamma_p (\omega) 
 &\equiv& {\pi \over V} \sum_{\vec{k}} ~\left|V_p (\vec{k}) \right|^2 ~
   \delta (\hbar\omega - \epsilon_{p\vec{k}} ).
\end{eqnarray}
\end{mathletters}
The above equation for the current is quite general and can be used as a starting 
point in reducing the expression of current in specific cases.

The average number of electrons at a side-connected dot $S$ should remain the same 
in a steady state. This condition is essential to get the right expression for
the current flowing from the left to the right electrodes. That is, the net current flowing 
out of the dot $S$ should vanish for
\begin{eqnarray} 
I_S &=& {e\over i\hbar} \sum_{\alpha} \left[ 
     W^* < d_{S\alpha}^{\dag} d_{A\alpha}^{\phantom{*}} > 
    - W < d_{A\alpha}^{\dag} d_{S\alpha}^{\phantom{*}} > \right] 
 ~=~ -ie \sum_{\alpha} 
   \left[ W^* G_{AS}^{<} (t,t) - W G_{SA}^{<} (t,t) \right]. 
\end{eqnarray}
The current $I_S$ was expressed in terms of the mixed Green's functions
\begin{mathletters}
\begin{eqnarray}
i\hbar G_{AS} (t,t') 
 &=& <T d_{A\alpha}^{\phantom{*}} (t)  d_{S\alpha}^{\dag} (t') >, \\
i\hbar G_{SA} (t,t') 
 &=& <T d_{S\alpha}^{\phantom{*}} (t)  d_{A\alpha}^{\dag} (t') >.
\end{eqnarray}
\end{mathletters}
These mixed Green's functions be written as
\begin{mathletters}
\begin{eqnarray}
G_{AS} (t,t') 
 &=& W \int_C dt_1 ~ \tilde{D}_{A}(t,t_1) D_S(t_1, t') 
 ~=~ W \int_C dt_1 ~ D_A(t,t_1) \tilde{D}_S(t_1, t'), \\
G_{SA} (t,t') 
 &=& W^* \int_C dt_1 ~ \tilde{D}_S(t,t_1) D_A(t_1, t') 
 ~=~ W^* \int_C dt_1 ~ D_S(t,t_1) \tilde{D}_A(t_1, t').
\end{eqnarray}
\end{mathletters}
Here $D_A$ and $D_S$ are the fully dressed Green's functions of the dots $A$ and $S$, 
respectively. 
\begin{mathletters}
\begin{eqnarray}
i\hbar D_A (t,t') 
 &=& <T d_{A\alpha}^{\phantom{*}} (t)  d_{A\alpha}^{\dag} (t') >, \\
i\hbar D_S (t,t') 
 &=& <T d_{S\alpha}^{\phantom{*}} (t)  d_{S\alpha}^{\dag} (t') >. 
\end{eqnarray}
\end{mathletters}
On the other hand $\tilde{D}_A$ and $\tilde{D}_S$ are the corresponding 
Green's functions which cannot be separated into two parts by removing one 
tunneling matrix $W$. Using analytic continuation the current $I_S$ can 
be written as
\begin{eqnarray}
I_S &=& -ie |W|^2 \sum_{\alpha} \int {d\omega \over 2\pi} 
  \left\{ [ \tilde{D}_A^{r}(\omega) - \tilde{D}_A^{a}(\omega) ] D_S^{<}(\omega) 
        - [ D_S^{r}(\omega) - D_S^{a}(\omega) ] \tilde{D}_A^{<}(\omega) \right\} 
      \nonumber\\
 &=& e \sum_{\alpha} \int {d\omega \over 2\pi} ~
   2|W|^2 | \tilde{D}_A^{r} (\omega) |^2 \left[ 
      \pi \tilde{\Sigma}_A^{<} (\omega) A_S(\omega) 
     + \mbox{Im}\{ \tilde{\Sigma}_A^{r} (\omega) \} D_S^{<} (\omega) \right]. 
\end{eqnarray}
Since $I_S = 0$ in steady state, we find the equation relating the lesser part to 
the retarded part of the Green's function or the spectral function $A_S$
for the dot $S$.
\begin{mathletters}
\begin{eqnarray}
D_S^{<} (\omega)
 &=& \pi { \tilde{\Sigma}_A^{<} (\omega) \over 
           - \mbox{Im} \tilde{\Sigma}_A^{r} (\omega) } A_S(\omega), \\
f_{\rm eff} (\omega) 
 &=& { \tilde{\Sigma}_A^{<} (\omega) \over 
           - 2\mbox{Im} \tilde{\Sigma}_A^{r} (\omega) }.
\end{eqnarray}
\end{mathletters}
That is, the condition that no net current flows out of the dot $S$ determines
its nonequilibrium thermal distribution function. The current $I_S$ can be 
expressed in another form
\begin{eqnarray}
I_S &=& -ie |W|^2 \sum_{\alpha} \int {d\omega \over 2\pi} 
  \left\{ [ D_A^{r}(\omega) - D_A^{a}(\omega) ] \tilde{D}_S^{<}(\omega) 
        - [ \tilde{D}_S^{r}(\omega) - \tilde{D}_S^{a}(\omega) ] D_A^{<}(\omega) 
  \right\} \nonumber\\
 &=& e \sum_{\alpha} \int {d\omega \over 2\pi} ~
   2|W|^2 |D_A^{r} (\omega) |^2 \left[ 
      \pi \Sigma_A^{<} (\omega) \tilde{A}_S(\omega) 
     + \mbox{Im}\{ \Sigma_A^{r} (\omega) \} \tilde{D}_S^{<} (\omega) \right]. 
\end{eqnarray}

\subsection{Interacting Case: $U_A = 0$ and $U_S \neq 0$}
 When $U_A = 0$ and $U_S \neq 0$, the active dot $A$ becomes a resonant level 
and the side-connected dot $S$ is strongly correlated. 
This model Hamiltonian may describe the situation that 
the dot $A$ lies in the conductance peak region and the dot
$S$ in the Kondo or Coulomb blockade region. 
Green's functions of the two dots are given by the 
Dyson equations with the self-energies being given by 
\begin{mathletters}
\begin{eqnarray}
\Sigma_A (t,t') &=& \Sigma_c (t,t') + W^2 \tilde{D}_S (t,t'), \\
\Sigma_S (t,t') &=& \Sigma_U (t,t') + W^2 \tilde{D}_A (t,t').
\end{eqnarray}
\end{mathletters}
Here $\Sigma_c(t,t') = \Sigma_{cL}(t,t') + \Sigma_{cR}(t,t')$ 
[equation (\ref{csigma})] 
is the self-energy coming from tunneling into left and right electrodes, 
$\Sigma_U(t,t')$ is the self-energy for a dot $S$ due to the 
on-site Coulomb interaction, and the auxillary Green's functions are 
defined by the Dyson equations
\begin{mathletters}
\begin{eqnarray}
\label{auxgreen}
\tilde{D}_A (t,t') &=& D_{A0} (t,t') 
  + \int_C dt_1 \int_C dt_2 ~\tilde{D}_A (t,t_1) \Sigma_c(t_1,t_2) D_{A0}(t_2,t'), \\
\tilde{D}_S (t,t') &=& D_{S0} (t,t') 
  + \int_C dt_1 \int_C dt_2 ~\tilde{D}_S (t,t_1) \Sigma_U(t_1,t_2) D_{S0}(t_2,t'). 
\end{eqnarray}
\end{mathletters}
From the self-energy equations we find 
\begin{mathletters}
\begin{eqnarray}
\Sigma_A^{<}(\omega) 
 &=& 2\Gamma_L(\omega) f_L(\omega) + 2\Gamma_R(\omega) f_R(\omega)
    + W^2 \tilde{D}_{S}^{<} (\omega), \\
\Sigma_A^{r}(\omega) - \Sigma_A^{a}(\omega)
 &=& -2i [ \Gamma_L(\omega) + \Gamma_R(\omega) ] 
    + |W|^2 \left[ \tilde{D}_{S}^{r} (\omega) - \tilde{D}_{S}^{a} (\omega) \right].  
\end{eqnarray}
\end{mathletters}
Inserting all the equations into the expression of the current (\ref{currenteqn}), 
we find
\begin{eqnarray}
I_p &=& e \sum_{\alpha} \int{d\omega \over 2\pi} 
  \left\{  
   4\Gamma_L(\omega) \Gamma_R(\omega) |D^r(\omega)|^2 
     [ f_{\bar{p}} (\omega) - f_p(\omega) ] 
  + 2 |W|^2 \Gamma_p(\omega) |D_{A}^r(\omega)|^2 
     [ \tilde{D}_{S}^{<} (\omega) - 2\pi f_p(\omega) \tilde{A}_S (\omega) ] 
  \right\}
\end{eqnarray}
From the condition $I_S=0$, we have
\begin{mathletters}
\begin{eqnarray}
\tilde{D}_{S}^{<} (\omega)
 &=& \pi {\Sigma_A^{<} (\omega) \over - \mbox{Im} \Sigma_A^r (\omega) } 
   \tilde{A}_S (\omega) \nonumber\\
 &=& { 2\pi [ \Gamma_L(\omega) f_L(\omega) + \Gamma_R(\omega) f_R(\omega) ] 
       + \pi |W|^2 \tilde{A}_S^{<} (\omega) 
     \over \Gamma_L(\omega) + \Gamma_R(\omega) + \pi W||^2 \tilde{A}_S (\omega) } 
   \times \tilde{A}_S (\omega) \nonumber\\
 &=& 2\pi f_{\rm eff} (\omega) \tilde{A}_S (\omega), \\
f_{\rm eff} (\omega)
 &\equiv& { \Gamma_L(\omega) f_L(\omega) + \Gamma_R(\omega) f_R(\omega)
     \over \Gamma_L(\omega) + \Gamma_R(\omega) }. \nonumber
\end{eqnarray}
\end{mathletters}
It follows from the relation $I_S=0$ that $D_{S}^{<} (\omega) 
= 2\pi f_{\rm eff} (\omega) A_S(\omega)$.  
The effective Fermi-Dirac function $f_{\rm eff} (\omega)$ describes 
the nonequilibrium thermal distribution at the dot $S$ and also at the active dot $A$.
[It can be shown from the equation (\ref{greeneqn}).]
The current flowing from left to right electrode becomes
\begin{mathletters}
\begin{eqnarray}
\label{currenteqn2}
I_L &=& e \sum_{\alpha} \int{d\omega \over 2\pi} ~
   T(\omega) ~[ f_R (\omega - f_L (\omega) ], \\
T(\omega) 
 &=& 4\Gamma_L(\omega) \Gamma_R(\omega) |D_{A}^r(\omega)|^2 
   \left[ 1 + {\pi |W|^2 \tilde{A}_S(\omega) 
               \over \Gamma_L(\omega) + \Gamma_R(\omega)} \right] 
 ~=~ { 4\pi\Gamma_L(\omega) \Gamma_R(\omega) \over 
    \Gamma_L(\omega) + \Gamma_R(\omega) } A_A(\omega).
\end{eqnarray}
\end{mathletters}
Note that the current $I_L$ consists of two contributions, coming from 
the direct path ($L\to A \to R$) and the indirect path ($L \to A \to S \to A \to R$), 
which are interfereing each other. 

 The current is written down in terms of the transmission coefficient, that is, 
in Landauer-B\"{u}ttiker form. In turn the transmission coefficient is proportional 
to the spectral function of the active dot $A$. To find the spectral function
$A_A (\omega)$ we note that Green's functions of the two dots are related to 
each other by the equations 
\begin{eqnarray}
\label{greeneqn}
D_{A}^r (\omega) &=& \tilde{D}_{A}^r (\omega)
    + |W|^2 \tilde{D}_{A}^r (\omega) D_{S}^r (\omega) \tilde{D}_{A}^r (\omega).
\end{eqnarray}
To get the desired spectral function we have to calculate the Green's function 
$D_{S}^r$ at a side-connected dot and the auxillary Green's function 
$\tilde{D}_{A}^r$ at the active dot is given by the equation (\ref{auxgreen}). 
Due to the on-site Coulomb interaction at a side-connected dot $S$, 
the calculation of $D_{S}^{r} (\omega)$ is highly non-trivial 
and requires a many-body calculation. 
In the next section we describe the non-crossing approximation in order to find the 
Green's function at a side-connected dot $S$.

\section{Non-crossing Approximation (NCA)}
 To study the I-V curves of our model system, we need to calculate the 
Green's function at a side-connected dot $S$ out of equilibrium. 
For this purpose we adopt the NCA which has been 
a very fruitful tool for the study of Anderson model in
both equilibrium\cite{eqnca1,eqnca2} and nonequilibrium\cite{neqnca1,neqnca2}.

To begin with, we need to identify the 
effective continuum band for the dot $S$. In the Anderson model which describes
the magnetic ions imbedded in normal metals, the self-energy 
can be written as a sum of two contributions.
\begin{eqnarray}
\label{anderson}
\Sigma_f (\omega) 
 &=& \int {d\epsilon \over \pi} {\Gamma_{\rm eff} (\epsilon) 
   \over \hbar \omega - \epsilon + i\delta} + \Sigma_U^{r} (\omega).
\end{eqnarray}
The first term is the one-body contribution from tunneling into the conduction band
and the second term $\Sigma_U$ is the many-body contribution from 
the on-site Coulomb interaction. 
On the other hand the self-energy of the dot $S$ in our model is given by 
\begin{eqnarray}
\label{ourmodel}
\Sigma_S (\omega) 
 &=& |W|^2 \tilde{D}_{A}^{r} (\omega) + \Sigma_U^{r} (\omega). 
\end{eqnarray}
Comparing the two self-energies (equations \ref{anderson} and \ref{ourmodel})
we can identify that the effective Anderson hybridization function
is given by the equation
\begin{eqnarray}
\Gamma_{\rm eff} (\epsilon)
 &=& - |W|^2 \mbox{Im} \tilde{D}_{A}^{r} (\epsilon)
 ~=~ - |W|^2 \mbox{Im} \left\{ {1\over \epsilon - E_d - \Sigma_c^{r} (\epsilon) }
   \right\} ~=~ \pi |W|^2 \tilde{A}_A (\epsilon). 
\end{eqnarray}
The effective continuum band for the dot $S$ is represented by 
the auxillary Green's function $\tilde{D}_A(t,t')$. 

\begin{figure}
\protect\centerline{\epsfxsize=3.0in \epsfbox{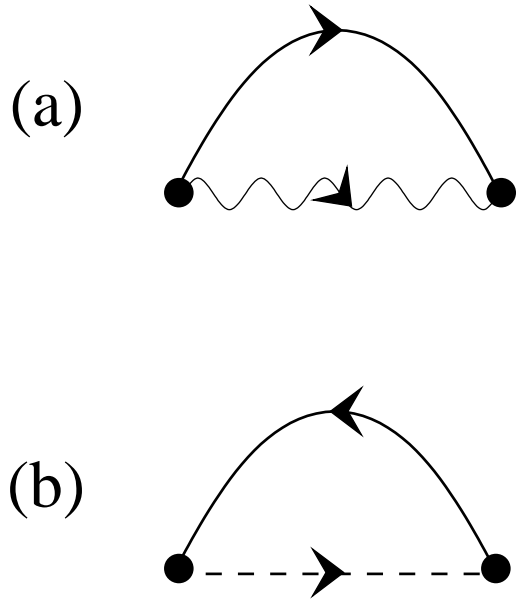}}
\vskip 0.5cm
\protect\caption[Fig.1.]
{NCA self-energy diagrams for (a) pseudo fermion and (b) slave boson. 
The solid line represents the propagator $\tilde{D}$, 
the Green's function of the active dot $A$ when $W=0$. The dashed (wavy) line 
is the pseudo fermion (boson) propagator. The solid circle means the 
tunneling matrix $W$.}
\protect\label{nca}
\end{figure}

To find the Green's function
for the dot $S$ we do a perturbative expansion in $W$. 
To begin with we represent the Fock space of the dot $S$ by the pseudo 
particle operators. In the large $U_S$ limit we may neglect the double 
occupation. 
\begin{eqnarray}
d_{S\alpha} &\to& b^{\dag} f_{\alpha}^{\phantom{*}}. 
\end{eqnarray}
The boson operator $b$ annihilates the empty state and the pseudo fermion 
operator $f_{\alpha}$ destroys the singly occupied state at the dot $S$.
After replacing the electron operator $d_{S\alpha}$ by the pseudo particle operators,
the occupation constraint has to be enforced: $Q=1$.
\begin{eqnarray}
Q &=& \sum_{\alpha} f_{\alpha}^{\dag} f_{\alpha}^{\phantom{*}} + b^{\dag}b.
\end{eqnarray}
This constraint is taken into account using the Lagrange multiplier 
method: $H \to H + \lambda(Q-1)$. The projection to the physical Hilbert 
space $Q=1$ leads to the following projection rule for the physical observables:
\begin{eqnarray}
O(\omega) 
 &=& {1\over Z_S} \lim_{\lambda \to \infty} e^{\beta\lambda} 
    O_{\lambda} (\omega \to \omega+\lambda), \\
Z_S &=& \lim_{\lambda \to \infty} e^{\beta\lambda}  <Q>_{\lambda}. 
\end{eqnarray}
Here $O_{\lambda} (\omega)$ is a thermal average of observable operator 
$\hat{O}$ for the Hamiltonian $H + \lambda(Q-1)$.
The Feynman diagram rules for the Green's functions are very simple. 
In the $2n$-th order there are $3n+1$ pairs of creation and annihilation operators. 
This leads to the factor: $[i\hbar]^{-2n} [i\hbar]^{3n+1}$. Canceling one factor
$i\hbar$ leads to the following factor: $(-1)^{N_F} [i\hbar]^n |W|^{2n}$ where 
$N_F$ is the number of closed fermion loops. With this rule the NCA self-energies
(displayed in Fig.~\ref{nca}) are given by the equations
\begin{mathletters}
\begin{eqnarray}
\Sigma_f (\lambda; t,t') 
 &=& i\hbar |W|^2 ~G_b(\lambda; t,t') \tilde{D}_{A} (t,t'), \\
\Sigma_b (\lambda; t,t') 
 &=& -i\hbar N_s |W|^2 ~G_f(\lambda; t,t') \tilde{D}_{A} (t',t). 
\end{eqnarray}
\end{mathletters}
Here $N_s=2$ accounts for the spin degeneracy of singly occupied state for the dot 
$S$. $G_b$ and $G_f$ are the pseudo particle Green's functions of boson and 
fermion, respectively. Using analytic continuation to the real-time axis 
\cite{langreth} we find
\begin{mathletters}
\begin{eqnarray}
\Sigma_f^{>} (\lambda; t,t')
 &=& |W|^2 G_b^{>} (\lambda; t,t') \tilde{D}_{A}^{>} (t,t'), \\
\Sigma_f^{<} (\lambda; t,t')
 &=& - |W|^2 G_b^{<} (\lambda; t,t') \tilde{D}_{A}^{<} (t,t'), \\
\Sigma_f^{r} (\lambda; t,t')
 &=& |W|^2 \left[ G_b^{r} (\lambda; t,t') \tilde{D}_{A}^{>} (t,t')
       - G_b^{<} (\lambda; t,t') \tilde{D}_{A}^{r} (t,t') \right], \\
\Sigma_b^{>} (\lambda; t,t')
 &=& N_s |W|^2 G_f^{>} (\lambda; t,t') \tilde{D}_{A}^{<} (t',t), \\
\Sigma_b^{<} (\lambda; t,t')
 &=& - N_s |W|^2 G_f^{<} (\lambda; t,t') \tilde{D}_{A}^{>} (t',t), \\
\Sigma_b^{r} (\lambda; t,t')
 &=& N_s |W|^2 \left[ G_f^{r} (\lambda; t,t') \tilde{D}_{A}^{<} (t',t)
       + G_f^{<} (\lambda; t,t') \tilde{D}_{A}^{a} (t',t) \right]. 
\end{eqnarray}
\end{mathletters}
To give a concrete example of analytic continuation we consider the derivation of
$\Sigma_b^{>} (\lambda; t,t')$: 
\begin{eqnarray}
{1\over i} \Sigma_b^{>} (\lambda; t,t')
 &=& -i\hbar |W|^2 ~{1\over i} G_f^{>}(\lambda; t,t') \times
       {-1\over i} \tilde{D}_{A}^{<}(t',t).
\end{eqnarray}
Removing the common factors on both sides of the equation, we find the desired
expression for $\Sigma_b^{>} (\lambda; t,t')$. 

 After Fourier transform, we get the self-energy equations
\begin{mathletters}
\begin{eqnarray}
\Sigma_f^{>}(\lambda; \omega) 
 &=& |W|^2 \int{d\zeta\over 2\pi}~ G_b^{>} (\lambda; \omega-\zeta) 
    \tilde{D}_{A}^{>} (\zeta), \\
\Sigma_f^{<}(\lambda; \omega) 
 &=& - |W|^2 \int{d\zeta\over 2\pi}~ G_b^{<} (\lambda; \omega-\zeta) 
    \tilde{D}_{A}^{<} (\zeta), \\
\Sigma_f^{r}(\lambda; \omega) 
 &=& |W|^2 \int{d\zeta\over 2\pi}~ \left[ 
     G_b^{r} (\lambda; \omega-\zeta) \tilde{D}_{A}^{>} (\zeta)
       - G_b^{<} (\lambda; \omega-\zeta) \tilde{D}_{A}^{r} (\zeta) \right], \\    
\Sigma_b^{>}(\lambda; \omega) 
 &=& N_s |W|^2 \int{d\zeta\over 2\pi}~ G_f^{>} (\lambda; \omega+\zeta) 
    \tilde{D}_{A}^{<} (\zeta), \\
\Sigma_b^{<}(\lambda; \omega) 
 &=& - N_s |W|^2 \int{d\zeta\over 2\pi}~ G_f^{<} (\lambda; \omega+\zeta) 
    \tilde{D}_{A}^{>} (\zeta),  \\
\Sigma_b^{r}(\lambda; \omega) 
 &=& N_s|W|^2 \int{d\zeta\over 2\pi}~ \left[ 
      G_f^{r} (\lambda; \omega+\zeta) \tilde{D}_{A}^{<} (\zeta)
       + G_f^{<} (\lambda; \omega+\zeta) \tilde{D}_{A}^{a} (\zeta) \right].   
\end{eqnarray}
\end{mathletters}
The Green's function at the dot $S$ can be written down in terms of pseudo 
particle operators.
\begin{eqnarray}
i\hbar D_S(\lambda; t,t') 
 &=& <T d_{S\alpha}^{\phantom{*}} (t) d_{S\alpha}^{\dag} (t') > 
 ~=~ <T b^{\dag}(t) f_{\alpha}^{\phantom{*}} (t) f_{\alpha}^{\dag} (t') b(t') >.
\end{eqnarray}
In the leading approximation the desired Green's function is 
\begin{eqnarray}
D_S (\lambda; t,t') &\approx& i\hbar ~ G_f(\lambda; t,t') G_b(\lambda; t',t).
\end{eqnarray}
Analytic continuation to the real-time axis\cite{langreth} leads to 
\begin{mathletters}
\begin{eqnarray}
D_S^{>}(\lambda; t,t') 
 &=& - G_f^{>}(\lambda; t,t') G_b^{<}(\lambda; t',t), \\
D_S^{<}(\lambda; t,t') 
 &=& G_f^{<}(\lambda; t,t') G_b^{>}(\lambda; t',t), \\
D_S^{r}(\lambda; t,t') 
 &=& - G_f^{r}(\lambda; t,t') G_b^{<}(\lambda; t',t) 
      - G_f^{<}(\lambda; t,t') G_b^{a}(\lambda; t',t). 
\end{eqnarray}
\end{mathletters}
After Fourier transform we get
\begin{mathletters}
\begin{eqnarray}
D_S^{>}(\lambda; \omega) 
 &=& - \int{d\zeta \over 2\pi} ~ G_f^{>}(\lambda; \omega+\zeta) 
    G_b^{<}(\lambda; \zeta), \\
D_S^{<}(\lambda; \omega) 
 &=& \int{d\zeta \over 2\pi} ~G_f^{<}(\lambda; \omega+\zeta) 
    G_b^{>}(\lambda; \zeta), \\
D_S^{r}(\lambda; \omega) 
 &=& - \int{d\zeta \over 2\pi} ~\left[ 
        G_f^{r}(\lambda; \omega+\zeta) G_b^{<}(\lambda; \zeta)
         + G_f^{<}(\lambda; \omega+\zeta) G_b^{a}(\lambda; \zeta) \right]. 
\end{eqnarray}
\end{mathletters}
From the projection to the physical Hilbert space $Q=1$ of the Green's function
at the dot $S$, we find the projection rules for the pseudo particle Green's 
functions. 
\begin{eqnarray}
\label{sidegreen}
D_S^{r}(\omega) 
 &=& {1 \over Z_S} \lim_{\lambda\to\infty} 
    e^{\beta\lambda} D_S^{r}(\lambda; \omega) \nonumber\\
 &=& - {1 \over Z_S} \lim_{\lambda\to\infty} e^{\beta\lambda} 
    \int{d\zeta \over 2\pi} ~\left[ 
      G_f^{r}(\lambda; \omega+\zeta+\lambda) G_b^{<}(\lambda; \zeta+\lambda)
    + G_f^{<}(\lambda; \omega+\zeta+\lambda) G_b^{a}(\lambda; \zeta+\lambda) \right]
   \nonumber\\
 &=& - {1 \over Z_S} \int{d\zeta \over 2\pi} ~\left[ 
        G_f^{r}(\omega+\zeta) G_b^{<}(\zeta)
         + G_f^{<}(\omega+\zeta) G_b^{a}(\zeta) \right], \\
Z_S &=& \int {d\omega \over 2\pi} \left[ 
   N_s G_f^{<}(\omega) - G_b^{<}(\omega) \right].  
\end{eqnarray} 
Since the pseudo particle Green's functions are centered at $\omega=E_S+\lambda$ 
(pseudo fermion) and at $\omega=\lambda$ (slave boson), we have to shift the 
integration variable before taking the limit: $\lambda \to \infty$. 
For $D_S^r(\omega)$ to be well-defined the limiting pseudo particle Green's functions
should be defined as
\begin{mathletters}
\begin{eqnarray}
G_{b,f}^{r} (\omega) 
 &\equiv& \lim_{\lambda\to\infty} G_{b,f}^{r} (\lambda; \omega+\lambda), \\
G_{b,f}^{<}(\omega) 
 &\equiv& \lim_{\lambda\to\infty} e^{\beta\lambda} 
    G_{b,f}^{<} (\lambda; \omega+\lambda).
\end{eqnarray}
\end{mathletters}
The second equation means that $\lim_{\lambda\to\infty} 
G_{b,f}^{<} (\lambda; \omega+\lambda) = 0$. Applying the projection procedure to
the lesser and greater Green's functions of the dot $S$ we find the additional 
projection rules for the pseudo particle Green's functions:
\begin{eqnarray}
G_{b,f}^{>}(\omega) 
 &\equiv& \lim_{\lambda\to\infty}
    G_{b,f}^{>} (\lambda; \omega+\lambda).
\end{eqnarray} 
Accordingly the lesser and greater Green's functions of the dot $S$ are given by 
the equations
\begin{mathletters}
\begin{eqnarray}
D_S^{<}(\omega) 
 &=& {1 \over Z_S} \int{d\zeta \over 2\pi} ~G_f^{<}(\omega+\zeta) G_b^{>}(\zeta), \\
D_S^{>}(\omega) 
 &=& - {1 \over Z_S} \int{d\zeta \over 2\pi} ~ G_f^{>}(\omega+\zeta) 
    G_b^{<}(\zeta).
\end{eqnarray}
\end{mathletters}
Applying the projection procedure to the physical Green's function $D_S$ 
at the dot $S$ we obtained the projection rules for the pseudo particle 
Green's functions. 

 We are now in a position to find the self-energy equations projected to the 
physical Hilbert space, $Q=1$, for the pseudo particle Green's functions. 
\begin{mathletters}
\begin{eqnarray}
\Sigma_{b,f}^{r,>} (\omega)
 &=& \lim_{\lambda\to\infty} \Sigma_{b,f}^{r,>}(\lambda; \omega+\lambda), \\ 
\Sigma_{b,f}^{<} (\omega)
 &=& \lim_{\lambda\to\infty} e^{\beta\lambda} 
   \Sigma_{b,f}^{<}(\lambda; \omega+\lambda). 
\end{eqnarray}
\end{mathletters}
By comparing the self-energies we can see the identities: 
$G_{b,f}^{r}(\omega) = G_{b,f}^{>}(\omega)$.
\begin{mathletters}
\begin{eqnarray}
\Sigma_f^{<}(\omega) 
 &=& - |W|^2 \int{d\zeta\over 2\pi}~ G_b^{<} (\omega-\zeta) 
    \tilde{D}_{A}^{<} (\zeta), \\
\Sigma_f^{r}(\omega) 
 &=& |W|^2 \int{d\zeta\over 2\pi}~ G_b^{r} (\omega-\zeta) 
    \tilde{D}_{A}^{>} (\zeta), \\    
\Sigma_b^{<}(\omega) 
 &=& - N_s |W|^2 \int{d\zeta\over 2\pi}~ G_f^{<} (\omega+\zeta) 
  \tilde{D}_{A}^{>} (\zeta), \\
\Sigma_b^{r}(\omega) 
 &=& N_s|W|^2 \int{d\zeta\over 2\pi}~ G_f^{r} (\omega+\zeta) \tilde{D}_{A}^{<} (\zeta).   
\end{eqnarray}
\end{mathletters}
These are the NCA self-energy equations out of equilibrium.
Now inserting the following equations
\begin{mathletters}
\begin{eqnarray}
\tilde{D}_{A}^{<} (\omega) 
 &=& |\tilde{D}_{A}^r (\omega)|^2 ~[ 2\Gamma_L(\omega) f_L(\omega) 
     + 2\Gamma_R(\omega) f_R(\omega) ], \\
\tilde{D}_{A}^{>} (\omega) 
 &=& |\tilde{D}_{A}^r (\omega)|^2 ~[ 2\Gamma_L(\omega) f_L(-\omega) 
     + 2\Gamma_R(\omega) f_R(-\omega) ], 
\end{eqnarray} 
\end{mathletters}
we find the final version of the NCA equations
\begin{mathletters}
\begin{eqnarray}
\Sigma_f^{<}(\omega) 
 &=& - \sum_{p=L,R} \int{d\zeta\over \pi}~ |W|^2 |\tilde{D}_{A}^r(\zeta)|^2 
    \Gamma_p(\zeta) ~f_p(\zeta)~ G_b^{<} (\omega-\zeta), \\
\Sigma_f^{r}(\omega) 
 &=& \sum_{p=L,R} \int{d\zeta\over \pi}~  |W|^2 |\tilde{D}_{A}^r(\zeta)|^2 
    \Gamma_p(\zeta) ~f_p(-\zeta)~ G_b^{r} (\omega-\zeta), \\    
\Sigma_b^{<}(\omega) 
 &=& - N_s \sum_{p=L,R} \int{d\zeta\over \pi}~ |W|^2 |\tilde{D}_{A}^r(\zeta)|^2 
    \Gamma_p(\zeta) ~f_p(-\zeta)~ G_f^{<} (\omega+\zeta), \\
\Sigma_b^{r}(\omega) 
 &=& N_s \sum_{p=L,R} \int{d\zeta\over \pi}~  |W|^2 |\tilde{D}_{A}^r(\zeta)|^2 
    \Gamma_p(\zeta) ~f_p(\zeta)~ G_f^{r} (\omega+\zeta).
\end{eqnarray}
\end{mathletters}
The effective Anderson hybridization for a side-connected dot $S$ is given by 
$\Gamma_{\rm eff} (\omega) = |W|^2 |\tilde{D}_{A}^r(\omega)|^2 
[\Gamma_L(\omega) + \Gamma_R(\omega)]$. 
Although the dot $S$ is not directly coupled to the external electrodes
it can experience the bias potential difference via the active dot $A$.

\section{Numerical Results}
 For the numerical calculations, we assume the constant energy-independent 
tunneling matrix between the active quantum dot $A$ and two electrodes. 
The left and right electrodes are assumed to be described by the same Lorentzian 
density of states (DOS). The self-energies due to the tunneling into the electrodes 
are found in a closed form in this case. The auxillary Green's function for 
the active dot $A$ ($\tilde{D}_{A}^r$) and the effective Anderson hybridization 
function for the side-connected dot $S$
($\Gamma_{\rm eff} (\omega) = \tilde{\Gamma}_L(\omega) 
 + \tilde{\Gamma}_R(\omega)$) are given by the equations
\begin{eqnarray}
\tilde{D}_{A}^r (\omega) 
 &=& {1\over \hbar\omega - E_A - \Gamma \times 
        \displaystyle {D \over \hbar\omega + iD}}, \\
\tilde{\Gamma}_p (\omega) 
 &=& \Gamma_p |W|^2 ~ {D^2 \over [ \hbar\omega(\hbar\omega-E_A) - \Gamma D]^2 
     + (\hbar\omega-E_A)^2 D^2 }.
\end{eqnarray}
Here $D$ is the conduction band width of the Lorentzian DOS. 
The tunneling rate of an active dot $A$ into the electrodes
$\Gamma = \Gamma_L + \Gamma_R$ or the effective band width of Anderson hybridization
$\Gamma_{\rm eff} (\omega)$ will be taken 
as an energy unit in our numerical works. For this model the Kondo temperature 
at zero source-drain bias voltage can be estimated within an order of $1$ by 
the equation
\begin{eqnarray}
\label{kondot}
T_K &=& \Gamma \times \exp\left( - {\pi |E_S| \over 2\Gamma_{\rm eff}} \right).
\end{eqnarray}
Here $\Gamma_{\rm eff} = |W|^2 /\Gamma$ is the effective Anderson hybridization at
$\omega=0$ and $E_A=0$.

\begin{figure}
\epsfxsize=5.0in \noindent
\begin{minipage}[t]{.49\linewidth}
 \centering\epsfig{file=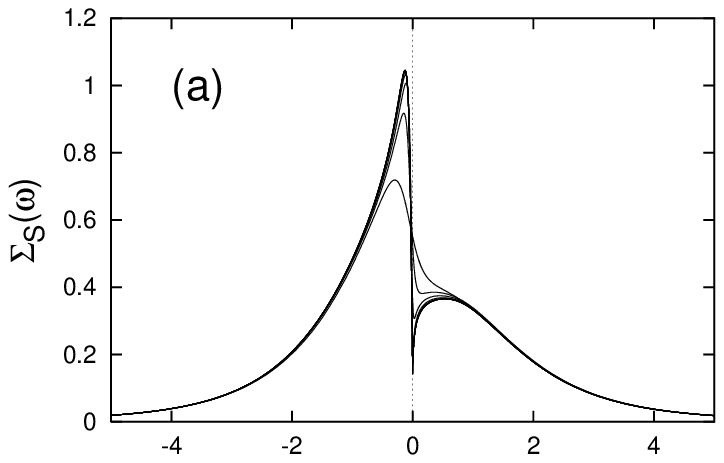,width=7.5cm}
\end{minipage}

\noindent
\begin{minipage}[t]{.49\linewidth}
 \centering\epsfig{file=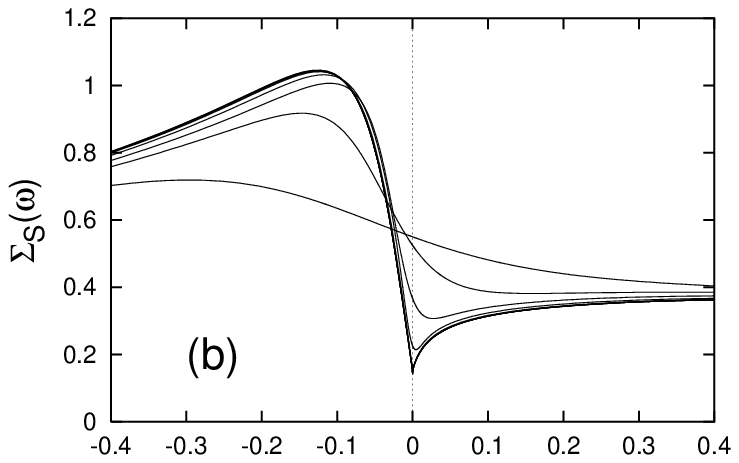,width=7.5cm}
\end{minipage}

\noindent
\begin{minipage}[t]{.49\linewidth}
 \centering\epsfig{file=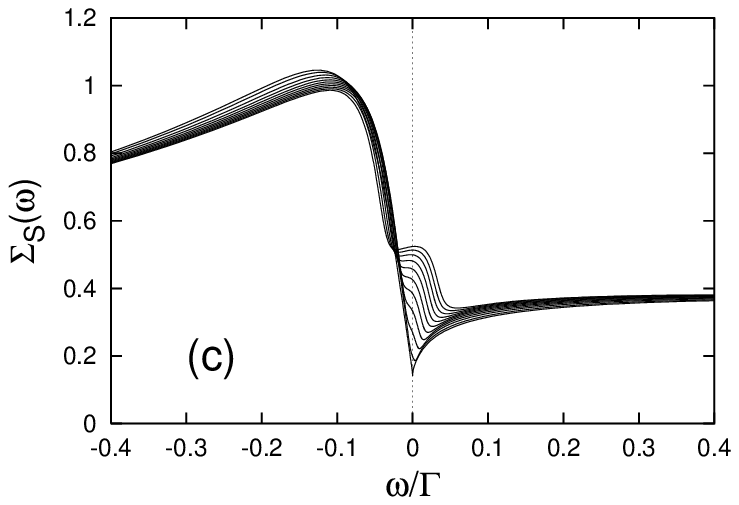,width=7.5cm}
\end{minipage}
\vspace{0.5cm} \caption{Temperature and bias voltage dependence of 
the self-energy of a side-connected dot $S$. Temperature variation of 
the self-energy is displayed in (a). The panel (b) is the magnified view 
of the panel (a) near $\omega=0$. With decreasing temperatures the structure 
in the self-energy becomes sharper near $\omega=0$. Temperature variations are
$T/T_K = 31.6, 10, 3.16, 1, 3.16\times 10^{-1}, 10^{-1}, 3.16\times 10^{-2}, 10^{-2},
3.16\times 10^{-3}, 10^{-3}$.
At $T= 10^{-3} \times T_K$ the variation of the self-energy with increasing bias voltage 
is displayed in (c). The source-drain bias voltages are varied from $eV =0$ to 
$eV = 20\times T_K$ by the amount $2\times T_K$.      
\label{self}}
\end{figure}

{\it Self-energies:} The self-energy $\Sigma_S$ of a side-connected dot $S$ 
is obtained from the Green's function $D_S$ which is calculated from 
equation (\ref{sidegreen}), using the relation 
$\Sigma_S^{r} = [D_S^{r}]^{-1} - [D_{S0}^{r}]^{-1}$. 
The imaginary part of $\Sigma_S^{r}$ is displayed in Fig~\ref{self}. 
Displayed results are corrected to satisfy the causality relation for the 
Green's function at the active dot $A$. This correction procedure is explained 
in the Appendix C and see also the references\cite{correction}.  
With decreasing temperature a sharp feature develops near 
$\omega=0$, which is a sign of Kondo resonance peak. When a finite source-to-drain
bias voltage is applied the sharp structure near $\omega=0$ gets smooth and 
split into two at $\omega = \pm eV/2$.

\begin{figure}
\epsfxsize=5.0in \noindent
\begin{minipage}[t]{.49\linewidth}
 \centering\epsfig{file=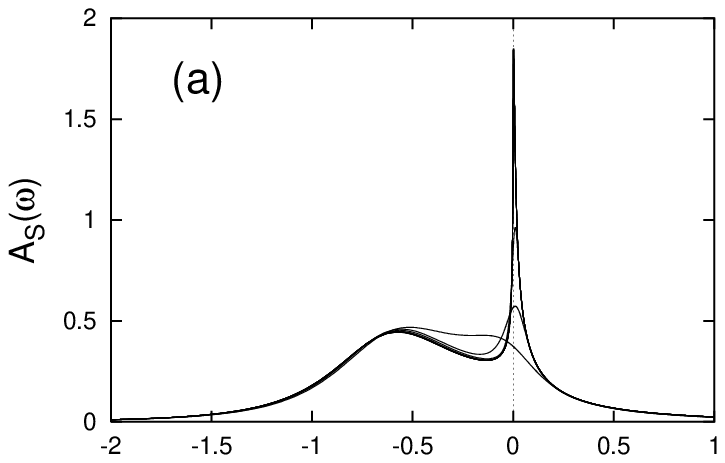,width=7.5cm}
\end{minipage}

\noindent
\begin{minipage}[t]{.49\linewidth}
 \centering\epsfig{file=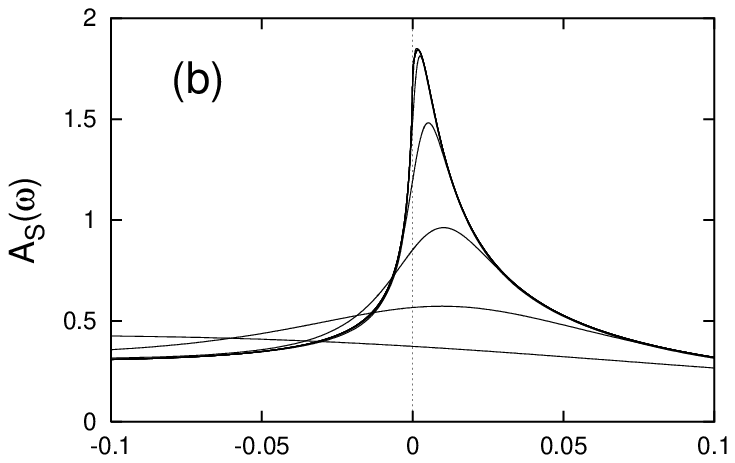,width=7.5cm}
\end{minipage}

\noindent
\begin{minipage}[t]{.49\linewidth}
 \centering\epsfig{file=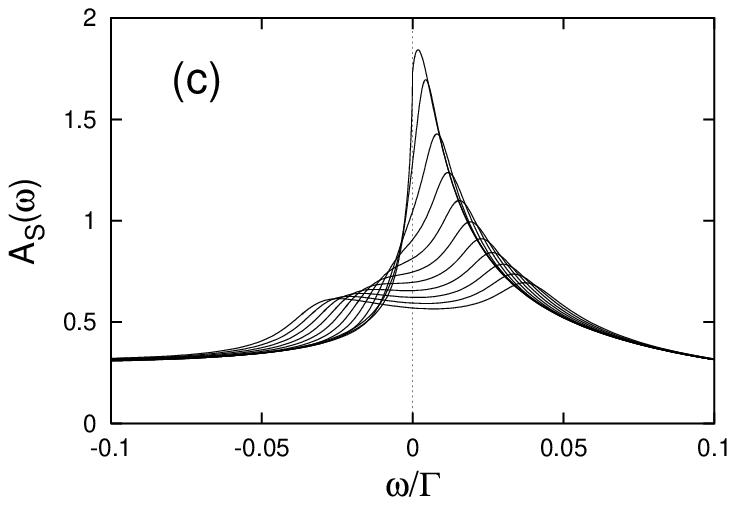,width=7.5cm}
\end{minipage}
\vspace{0.5cm} \caption{Temperature and bias voltage dependence of 
the spectral function $A_{S}(\omega)$ at a side-connected dot $S$. 
(a) and (b) show the developments of a Kondo resonance peak with 
decreasing temperature. The panel (b) is a magnified view 
of the panel (a). The bias voltage 
dependence of the spectral function $A_S(\omega)$ at $T= 10^{-3} \times T_K$ is 
displayed in (c). The Kondo resonance peak is progressively suppressed with 
the increasing bias voltage and develops into two-peak structure. 
Temperature and voltage variations are the same as in Fig.~\ref{self}.  
\label{specfftn}}
\end{figure}

\begin{figure}
\epsfxsize=5.0in \noindent
\begin{minipage}[t]{.49\linewidth}
 \centering\epsfig{file=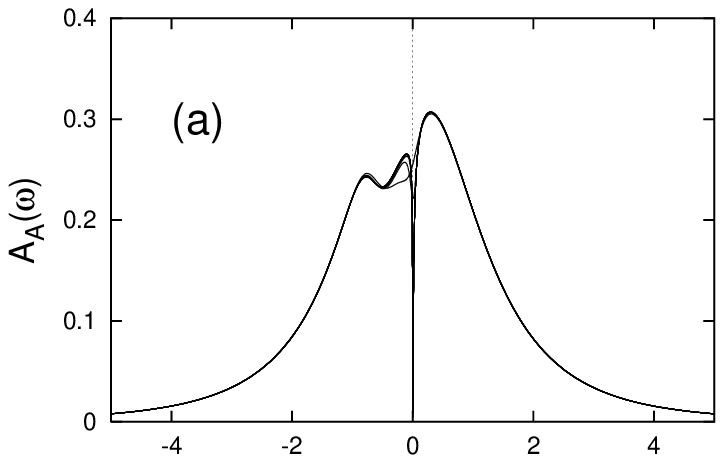,width=7.5cm}
\end{minipage}

\noindent
\begin{minipage}[t]{.49\linewidth}
 \centering\epsfig{file=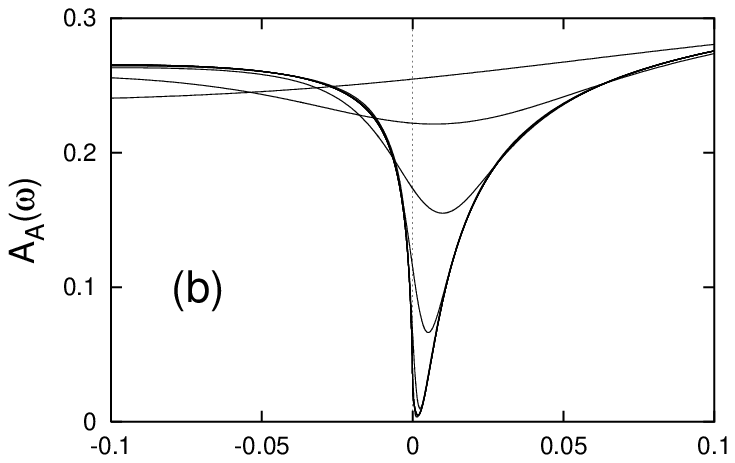,width=7.5cm}
\end{minipage}

\noindent
\begin{minipage}[t]{.49\linewidth}
 \centering\epsfig{file=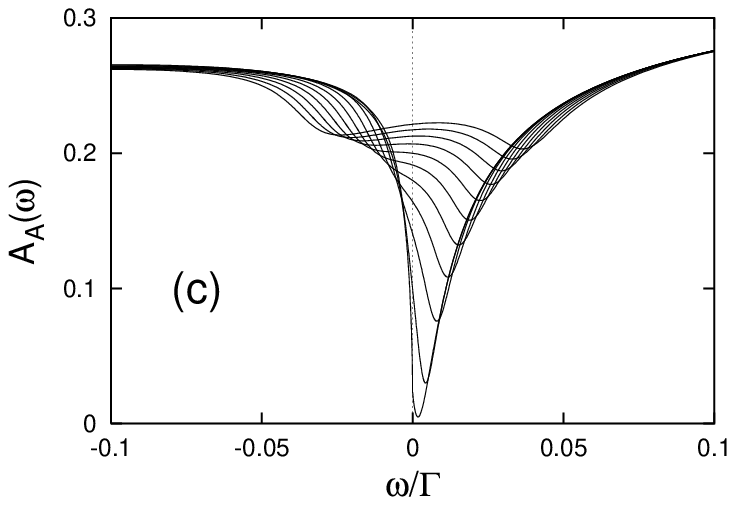,width=7.5cm}
\end{minipage}
\vspace{0.5cm} \caption{Temperature and bias voltage dependence of 
the spectral function $A_A(\omega)$ at the active dot $A$. 
Due to the Fano interference 
the spectral function $A_A (\omega)$ becomes depleted near $\omega=0$
as the temperature is lowered below $T_K$, as shown in (a) and (b). 
When a finite bias voltage is applied, the spectral depletion is refilled
and the two-peak structure in $A_S(\omega)$ is manifested as the double-dip
structure in $A_A(\omega)$.  
Temperature and voltage variations are the same as in Fig.~\ref{self}.  
\label{specdftn}}
\end{figure}

{\it Spectral functions:}
The spectral functions are calculated from the imaginary part of 
Green's functions 
\begin{eqnarray}
A_i (\omega) &=& - {1\over \pi} \mbox{Im} D_i^{r} (\omega), ~~i=A,S,
\end{eqnarray}
and displayed in Figs.~\ref{specfftn} and \ref{specdftn}. 
The spectral function $A_S(\omega)$ at a side-connected dot $S$ develops 
the Kondo resonance peak near $\omega=0$ with decreasing temperature below $T < T_K$. 
The Green's function at the active dot $A$ is renormalized according to 
the equation (\ref{greeneqn}). 
As the side-connected dot $S$ develops the Kondo resonance, the spectral function 
$A_A(\omega)$ at the active dot $A$ is 
depleted near $\omega=0$. The spectral depletion can be understood as the 
destructive interference between the direct path and the indirect path (Fano 
interference). 
The spectral depletion leads to the suppression of the 
current flowing through the dot $A$. The width of depleted region is given by the 
Kondo temperature (\ref{kondot}) of the dot $S$. When a finite value of 
source-to-drain bias voltage is applied, the peak in $A_S(\omega)$ and the dip 
in $A_A(\omega)$ near $\omega=0$ get splits into two at $\omega = \pm eV/2$.
The two Kondo resonance peaks in $A_S(\omega)$ and dips in $A_A(\omega)$
become pinned at the Fermi energies of the left and right electrodes.
\cite{neqnca1,neqnca2}  

 Since the NCA underestimates one-body contribution of the self-energy $\Sigma_S$
(side-connected dot) from tunneling into the effective continuum band, 
our NCA calculation of the 
spectral function at the active dot $A$ shows the negative spectral weight 
near the depleted region (violation of the causality relation). 
This is a purely artifact of the NCA. 
Since the NCA is based on the expansion in powers of the Anderson hybridization
(in our case $W$) and includes only a subset of diagrams up to the 
infinite order in $W$ without vertex corrections,
the underestimation is expected. Though the NCA
underestimates the one-body contribution of the self-energy, its dependence 
on the energy and temperature turns out to be correct close to 
$\omega=0$ and at $T>0$K. The NCA leads to the pathological results 
at and near $T=0$K.\cite{eqnca1} 
From the Fermi-liquid relation it can be shown that the spectral function 
of the active dot $A$ is zero at $\omega=0$ at zero temperature. 
According to the Fermi-liquid theory, the magnetic ion's spectral function
($A_S(\omega)$ in our case)  
can be related to the phase shift of conduction electrons 
(electrons at the active dot $A$ in our case) at the Fermi energy. 
The phase shift is determined by the 
Langreth-Friedel sum rule\cite{phase}, 
$n_S = 2{\delta\over \pi}$, where the numerical 
factor $2$ accounts for two possible spin directions.  
\begin{eqnarray}
A_S (0) &=& {1\over \pi \Gamma_{\rm eff} } \sin^2 {\pi n_S \over 2}
\end{eqnarray}
where $\Gamma_{\rm eff} = {|W|^2\over \Gamma}$ is the effective Anderson 
hybridization for a side-connected dot $S$ at $\omega=0$. 
With complete screening $n_S$ is equal to $1$. Inserting the Fermi-liquid relation 
into the equation (\ref{greeneqn}) it can be shown that the spectral function of the 
active dot $A$ vanishes at $\omega=0$, or $A_A(0) = 0$. 

 To compensate the NCA's underestimation of the self-energy we corrected 
the self-energy by adding a term proportional to one-body self-energy 
$|W|^2 \tilde{D}_A^{r} (\omega)$ to it such that 
the spectral function $A_A$ of the active dot remains non-negative definite 
and $A_A(0) \to 0$ with the temperature approaching zero. 
Since the contribution to the self-energy coming from the hybridization 
with the effective continuum band is of the one-body nature and 
temperature-independent, this correction proedure is not a bad idea.

\begin{figure}
\protect\centerline{\epsfxsize=4.0in \epsfbox{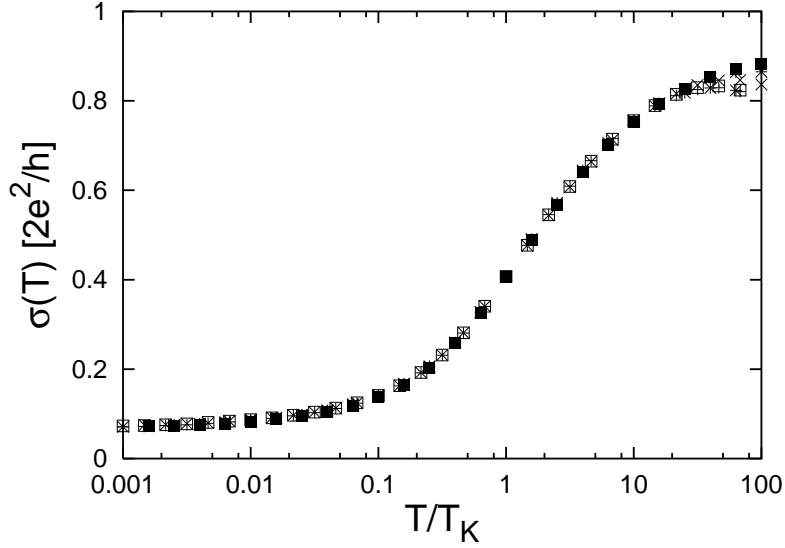}}
\vskip 0.5cm
\protect\caption[Fig.1.]
{Temperature dependence of linear response conductance $\sigma(T)$. 
The conductance at $V=0$ is suppressed with decreasing temperature 
as can be expected from 
the temperature dependence of the spectral function $A_A(\omega)$. 
The conductance is suppressed logarithmically due to the enhanced Kondo scattering
near the Kondo temperature and becomes saturated with decreasing temperature. 
$\sigma(T)$ shows the scaling behavior over a wide range of temperature. 
Symbols mean different sets of model parameters or differing Kondo temperatures.
}
\protect\label{eqG}
\end{figure}

{\it Linear response conductance:}
 The linear response conductance $\sigma(T)$ is calculated by differentiating 
the equation (\ref{currenteqn2}) with respect to the source-drain bias voltage 
at $V=0$. 
\begin{eqnarray}
\sigma(T) &=& {2e^2 \over h} \int d\epsilon ~T(\epsilon) ~
   \left[ - {\partial f(\epsilon) \over \partial \epsilon}\right].
\end{eqnarray}
 As shown in Fig.~\ref{eqG}, $\sigma(T)$ 
is suppressed with decreasing temperature. The conductance shows a logarithmic 
dependence on temperature near the Kondo temperature. 
From Fermi-liquid theory it can be expected
that the conductance is completely suppressed at zero temperature. 
The electron flow 
is blocked by the destructive inteference between two different paths: 
the direct path ($L\to A \to R$) and the indirect one 
($L \to A \to S \to A \to R$). 

 This suppressed behavior of $\sigma(T)$ at 
low temperatures is another example of zero-bias anomalies (ZBA) observed in several
mesoscopic devices. Scattering of conduction electrons off the Kondo impurities 
is one of scenarios explaining ZBA's. Depending on the topology of mesoscopic devices
and Kondo impurites, the ZBA can lead to the enhanced or suppressed conductance
at low temperature.  

 As expected from the Kondo effect, the linear response conductance $\sigma(T)$
shows the scaling behavior over a wide range of temperature. Several curves
of $\sigma(T)$ for different sets of model parameters (differing Kondo temperature)
collapse onto one curve as shown in Fig.~\ref{eqG}.

\begin{figure}
\protect\centerline{\epsfxsize=4.0in \epsfbox{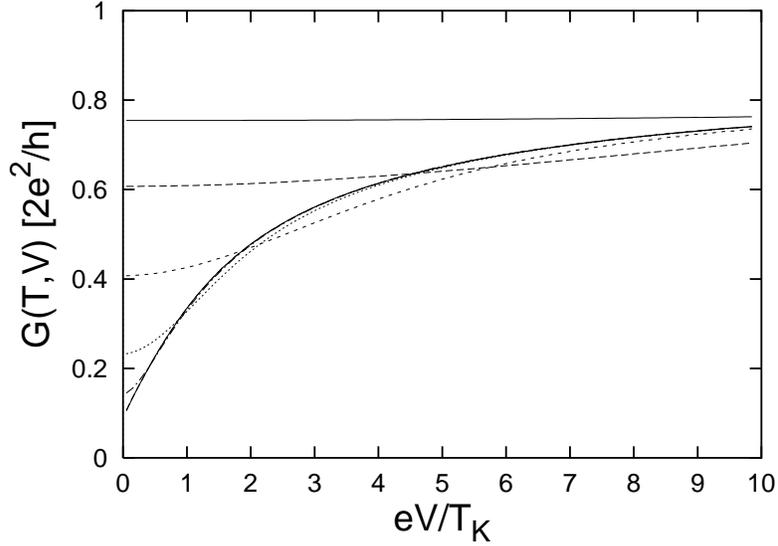}}
\vskip 0.5cm
\protect\caption[Fig.1.]
{Differential conductance $G(T,V) = dI/dV$ with varying temperature. The differential 
conductance is progressively suppressed near zero bias voltage with lowering
temperature. From top, the temperature variations are $T/T_K = 10, 3.16, 1, 
3.16\times 10^{-1}, 10^{-1}, 3.16\times 10^{-2}, 10^{-2}, 10^{-3}$.
The last three curves cannot be distinguished with the naked eye.}
\protect\label{neqI}
\end{figure}

{\it Differential conductance:}
 The current flowing from the left to right electrode is calculated using the 
equation (\ref{currenteqn2}). Differential conductance $G(T,V) = dI/dV$ 
is obtained by differentiating the current with respect to 
a finite source-drain bias voltage. 
 Differential conductance is displayed in Fig.~\ref{neqI}. With decreasing temperature
$dI/dV$ is suppressed near zero source-drain bias voltage. The pseudo gap behavior 
of the spectral function at the active quantum dot $A$ explains this $I-V$ 
characteristics.

\section{Conclusion}
 In this work we studied the transport properties of parallel double quantum dots
when only the {\it active} dot is coupled to the two external electrodes and 
the other dot is {\it side-connected} to the active dot. When the active
dot lies in the conductance peak region and the side-connected dot in the Coulomb 
blockade region, the current flow through the active dot is suppressed due to 
the Kondo scattering off the effective spin in the side-connected dot. 
Suppressed conductance at low temperature and near zero bias voltage 
can be understood in terms of Fano interference between two paths: direct 
path of left electrode - active dot - right electrode and 
indirect path of left electrode - active dot - side-coneccted dot
 - active-dot - right electrode. 
Our study suggests a possibility that the current 
flowing through one quantum dot can be controlled experimentally by an additional 
side-connected quantum dot.

\acknowledgments
 This work is supported in part by the BK21 project and in part by the National 
Science Foundation under Grant No. DMR 9357474.

\appendix

\section{Non-interacting Case: $U_A = U_S = 0$}
 When $U_A = U_S = 0$, both active and side-connected dots become resonant levels.
The current and its noise can be obtained analytically in a closed form. 
The self-energies of the dots $A$ and $S$ are 
\begin{eqnarray}
\Sigma_{A} (t,t') &=& \Sigma_c (t,t') + |W|^2 D_{S0} (t,t'), \\
\Sigma_{S} (t,t') &=& |W|^2 \tilde{D}_A (t,t').
\end{eqnarray}
Here the one-body self-energy $\Sigma_c$ is given by the equation
\begin{eqnarray}
\Sigma_c(t,t')
 &=& {1\over V} \sum_{p\vec{k}} |V_p(\vec{k})|^2 G_{cp}(\vec{k}; t,t'), \nonumber
\end{eqnarray}
and the auxillary Green's functions $\tilde{D}_A$ and $\tilde{D}_S$ 
are given by the equations
\begin{eqnarray}
\tilde{D}_{A} (t,t') 
 &=& D_{A0} (t,t') + \int_C dt_1 \int_C dt_2 ~ \tilde{D}_{A} (t,t_1) \Sigma_c(t_1, t_2) 
    D_{A0} (t_2,t'), \\
\tilde{D}_{S} (t,t') &=& D_{S0} (t,t').   
\end{eqnarray}
The self-energy of the auxillary Green's function $\tilde{D}_{A}$ is given by 
$\tilde{\Sigma}_{A} (t,t') = \Sigma_c(t, t')$. 

The retarded Green's functions can be readily obtained
\begin{eqnarray}
D_{A}^{r} (\omega)
 &=& { \hbar\omega - E_S \over [ \hbar\omega - E_S ] \times
       [ \hbar\omega - E_A  - \Sigma_c^{r} (\omega) ] - |W|^2 }, \\
D_{S}^{r} (\omega)
 &=& { \hbar\omega - E_A - \Sigma_c^{r} (\omega) \over [\hbar\omega - E_S] \times 
      [ \hbar\omega - E_A - \Sigma_c^{r} (\omega) ] - |W|^2 }.
\end{eqnarray}
Due to the Fano interference the spectral function of the active dot $A$ 
vanishes at $\omega = E_S$. On the other hand the spectral function of the 
side-connected dot $S$ is peaked at $\omega=E_S$. 
The nonequilibrium distribution function of the dot $S$ is also readily 
calculated
\begin{eqnarray}
\tilde{\Sigma}_{A}^{<} (\omega) 
 &=& 2\Gamma_L(\omega) f_L(\omega) + 2\Gamma_R(\omega) f_R(\omega), \nonumber\\
-\mbox{Im} \tilde{\Sigma}_{A}^{r} (\omega) 
 &=& \Gamma_L(\omega) + \Gamma_R(\omega), \nonumber\\
f_{\rm eff} (\omega) 
 &=& { \tilde{\Sigma}_{A}^{<} (\omega) 
       \over -2\mbox{Im} \tilde{\Sigma}_{A}^{r} (\omega) } 
 ~=~ { \Gamma_L(\omega) f_L(\omega) + \Gamma_R(\omega) f_R(\omega) 
      \over \Gamma_L(\omega) + \Gamma_R(\omega) }. 
\end{eqnarray}
The full self-energies of the active dot $A$ are 
\begin{eqnarray}
\Sigma_{A}^{<}(\omega) 
 &=& 2\Gamma_L(\omega) f_L(\omega) + 2\Gamma_R(\omega) f_R(\omega)
    + |W|^2 \tilde{D}_{S}^{<} (\omega), \nonumber\\
\Sigma_{A}^{r}(\omega) - \Sigma_{A}^{a}(\omega)
 &=& -2i [ \Gamma_L(\omega) + \Gamma_R(\omega) ] 
    + |W|^2 \left[ \tilde{D}_{S}^{r} (\omega) - \tilde{D}_{S}^{a} (\omega) \right].  
\end{eqnarray}
To determine $\tilde{D}_{S}^{<} (\omega)$, we use the condition 
$I_S = 0$ (the number of electrons in the dot $S$ remains the same). 
\begin{eqnarray}
\tilde{D}_{S}^{<} (\omega) 
 &=& 2\pi {\Gamma_L(\omega) f_L(\omega) + \Gamma_R(\omega) f_R(\omega) 
    \over \Gamma_L(\omega) + \Gamma_R(\omega)} \tilde{A}_S (\omega).
\end{eqnarray}
Here $\tilde{A}_S(\omega) = \delta(\hbar\omega - E_S)$. 

Inserting all the equations into the expression of the current (\ref{currenteqn}), 
we find  the current flowing from the left electrode to the right one.
\begin{eqnarray}
I_L &=& e \sum_{\alpha} \int{d\omega \over 2\pi} 
  \left\{  4\Gamma_L(\omega) \Gamma_R(\omega) |D_{A}^r(\omega)|^2 
     [ f_R (\omega) - f_L (\omega) ]  \right. \nonumber\\
 && \hspace{1.0cm} \left. 
   + 2 |W|^2 \Gamma_L (\omega) |D_{A}^r(\omega)|^2 
     [ \tilde{D}_{S}^{<} (\omega) - 2\pi f_L (\omega) \tilde{A}_S (\omega) ] 
  \right\}   \nonumber\\
 &=& {e\over h} \sum_{\alpha} \int d\epsilon ~
   T(\epsilon) [ f_R (\epsilon) - f_L (\epsilon) ].
\end{eqnarray}
The transmission spectral function $T(\epsilon)$ is given by the 
equation
\begin{eqnarray}
T(\epsilon) 
 &=& 4\Gamma_L(\epsilon) \Gamma_R(\epsilon) |D_{A}^r (\epsilon)|^2 
   \left[ 1 + {\pi |W|^2 \tilde{A}_S(\epsilon) 
               \over \Gamma_L(\epsilon) + \Gamma_R(\epsilon)} \right] 
 ~=~ { 4\pi\Gamma_L(\epsilon) \Gamma_R(\epsilon) \over 
    \Gamma_L(\epsilon) + \Gamma_R(\epsilon) } A_{A}(\epsilon).
\end{eqnarray}
The current noise, defined as the current-current correlation function, can be 
obtained in a closed form in the noninteracting case.
\begin{eqnarray}
S(t,t') &=& < \delta I(t) \delta I(t') > + < \delta I(t') \delta I(t) >. 
\end{eqnarray}
Here $\delta I = I - <I>$. Defining the current Green's function by
\begin{eqnarray}
i\hbar G_{II} (t,t') 
 &=& < T \delta I(t) \delta I(t') >, 
\end{eqnarray}
the current-current correlation function can be written as
$S(t,t') = \hbar G_{II}^{>} (t,t') + \hbar G_{II}^{>} (t',t)$. 
Six diagrams are contributing to the current noise and the static current 
noise $S_0 = S(\omega=0)$ is given by the well-known formula.
\cite{shotnoise}
\begin{eqnarray}
S_0 &=& {4e^2 \over h} \int d\epsilon ~ T(\epsilon) ~
   \left\{ f_L(\epsilon) [ 1 - f_L (\epsilon) ] 
        + f_R(\epsilon) [ 1 - f_R (\epsilon) ]  \right\} \nonumber\\
 && + {4e^2 \over h} \int d\epsilon ~ T(\epsilon) [ 1 - T(\epsilon) ] ~
     \left\{ f_L(\epsilon) - f_R (\epsilon) \right\}^2. 
\end{eqnarray}
The first line is the thermal Johnson noise and the second line is called 
the shot noise coming from a finite source-drain bias voltage.

\begin{figure}
\epsfxsize=5.0in \noindent
\begin{minipage}[t]{.49\linewidth}
 \centering\epsfig{file=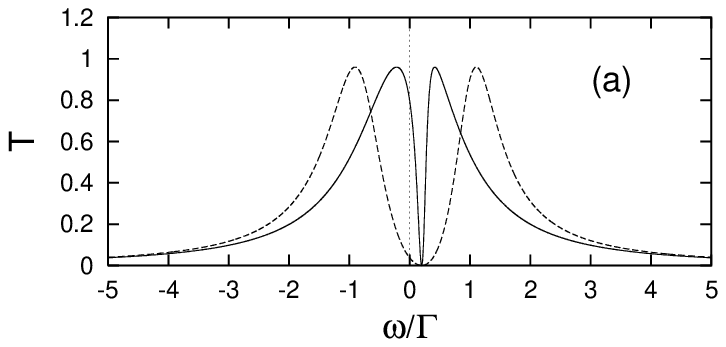,width=7.5cm}
\end{minipage}

\noindent
\begin{minipage}[t]{.49\linewidth}
 \centering\epsfig{file=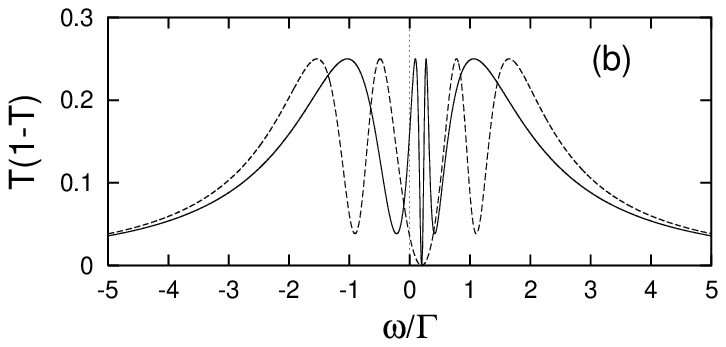,width=7.5cm}
\end{minipage}

\noindent
\begin{minipage}[t]{.49\linewidth}
 \centering\epsfig{file=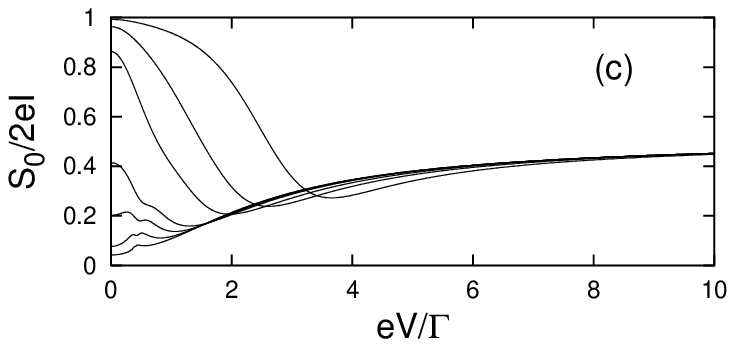,width=7.5cm}
\end{minipage}
\vskip 0.5cm
\protect\caption[Fig.1.]
{Tunneling coefficients and Fano factor at $T=0$K. Due to the Fano interference
the transmission coefficient $T(\omega)$ is completely suppressed when 
$\omega=E_S$, the energy level of the side-connected dot. The Fano factor,
$F = S_0/2eI$, has a dip structure accordingly. The model parameters are $E_A=0$,
$E_S/\Gamma=0.2$ for displayed curves. In (a) and (b), $W/\Gamma=0.3(1.0)$ for
solid(dashed) line, respectively. In panel (c), the values of $W/\Gamma$ are 
varied from the bottom successively as $0.1, 0.2, 0.3, 0.4, 0.7, 1.0, 1.5$.}
\protect\label{noise}
\end{figure}

 Typical results at $T=0$K are displayed in Fig.~\ref{noise} for 
the transmission coefficient $T(\epsilon)$ and the Fano factor $F= S_0/2eI$. 
Due to the Fano interference the transmission coefficient is 
completely suppressed at $\omega=E_S$, the energy level of the 
side-connected dot. This structure leads to a dip in the 
Fano factor. The Fano factor $F$ takes the following values at $V=0$ and 
$V\to\infty$. 
\begin{eqnarray}
F(V=0) &=& 1 - { 4\Gamma_L \Gamma_R \over (\Gamma_L + \Gamma_R)^2 } 
      {E_S^2 \over E_S^2 + (E_AE_S - W^2)^2 / \Gamma^2 }, \nonumber\\
F(V \to \infty) 
 &=& 1 - {2\Gamma_L \Gamma_R \over (\Gamma_L + \Gamma_R)^2}. \nonumber
\end{eqnarray}
Here $\Gamma = \Gamma_L + \Gamma_R$. The limiting value of $F$ at $eV \gg \Gamma$
is solely determined by the tunneling rates, $\Gamma_L$ and $\Gamma_R$.

\section{Interacting Case: $U_A \neq 0$ and $U_S \neq 0$}
 When both dots $A$ and $S$ are strongly correlated, the model system 
becomes highly non-trivial due to many-body effects. In this section
we derive the effective spin exchange interaction when both dots lie 
in the Kondo limit or in the absence of charge fluctuations.
Our model space ${\cal M}$ and its orthogonal space $\overline{\cal M}$ 
are then given by
\begin{eqnarray}
{\cal M} &=& |A\alpha> \otimes |S\beta> 
 ~=~ |A\alpha; S\beta>, ~~\alpha, \beta ~=~ \uparrow, \downarrow, \\
\overline{\cal M}
 &=& |A\alpha; S\rho>, ~ |A\rho; S\alpha>, ~ |A\rho; S\rho'>, ~~
  \alpha,\beta ~=~ \uparrow, \downarrow; ~~\rho,\rho' ~=~ \pm.  
\end{eqnarray}
Here $\alpha$ and $\beta$ represent the spin up or down state, while $\rho$ and $\rho'$ 
the isospin state:
\begin{eqnarray}
~|i+> &=& d_{i\uparrow}^{\dag} d_{i\downarrow}^{\dag} ~|0>, ~~~
 |i-> ~=~ |0>, ~~ i ~=~ A,S \nonumber
\end{eqnarray}
where $|0>$ means the empty state of a quantum dot. 

 To find the effective spin-exchange Hamiltonian we need to evaluate the following 
matrix elements
\begin{eqnarray}
< A\alpha; S\beta| H_1 | A\mu; S\rho> 
 &=& 0,  \nonumber\\
< A\alpha; S\beta| H_1 | A\rho; S\mu> 
 &=& \delta_{\beta,\mu} ~ <A\alpha| H_1 | A\rho> \nonumber\\
 &=& \delta_{\beta,\mu} \times \left\{ 
    \sum_{p\vec{k}} V_p^* (\vec{k}) c_{p\vec{k}\alpha}^{\phantom{*}} ~
      \delta_{\rho, -} 
    + (-1)^{\alpha-1/2} \sum_{p\vec{k}} V_p (\vec{k}) 
      c_{p\vec{k}\bar{\alpha}}^{\dag} ~\delta_{\rho, +}
   \right\} \nonumber\\
< A\alpha; S\beta| H_1 | A\rho; S\rho'>
 &=& (-1)^{\alpha-1/2} W~\delta_{\alpha+\beta,0} \delta_{\rho+\rho',0}. \nonumber  
\end{eqnarray} 
When the charge fluctuations can be neglected we find the effective spin-exchange
Hamiltonian
\begin{eqnarray}
\tilde{H}_1 
 &=& \sum_{p\vec{k}\alpha} \sum_{p'\vec{k}'\beta} 
   J_{pp'}(\vec{k},\vec{k}') ~c_{p\vec{k}\alpha}^{\dag}
     {1\over 2} \vec{\sigma}_{\alpha\beta} c_{p'\vec{k}'\beta}^{\phantom{*}}
     \cdot \vec{S}_A 
  + J_{AS} \left( \vec{S}_A \cdot \vec{S}_S - {1\over 4} \right), \\
J_{pp'}(\vec{k},\vec{k}')
 &=& 2 V_p(\vec{k}) V_{p'}^*(\vec{k}') \left[ {1\over -E_A} 
       + {1\over E_A + U_A} \right], \\
J_D &=& 2|W|^2 \left[ {1\over E_A + U_A - E_S} + {1\over E_S + U_S - E_A} \right].  
\end{eqnarray}
Electrons are scattered off an effective spin $\vec{S}_A$ at the active dot
and can flow from one electrode to the other. Since $\vec{S}_A$ is coupled 
antiferromagnetically to $\vec{S}_S$, the occurence of enhanced Kondo 
scattering at low temperature depends on the magnitude of antiferromagnetic 
coupling $J_{AS}$ relative to, e.g., the Kondo temperature. 
When $J_{AS} \gg T_K$, a spin-singlet formation between two dots is favored 
over the many-body Kondo singlet state and the enhanced Kondo scattering 
is not expected. On the other hand when $T_K \gg J_{AS}$, a Kondo singlet 
state is expected to be favored over a spin-singlet state between two dots 
and the conductance will be enhanced due to the Kondo scattering at 
low temperature. Detailed study of this model in equilibrium is in progress 
using the Wilson's numerical renormalization and will be published 
elsewhere.

\section{Fermi-Liquid Relations}
 In this section we discuss the correction of pathological behavior of NCA near 
$\omega=0$. The Green's function of the dot $A$ is related to that at the dot $S$ 
by the equation (\ref{greeneqn}). 
The causality is seldomly violated when the number of channels is larger than two.
On the other hand the causality is violated for the case of one-channel
near $\omega=0$ at temperatures below the Kondo temperature $T_K$. Recently 
developed conserving $t$-matrix approximation (CTMA)
\cite{ctma} may have a chance to overcome this pathology. In this paper 
we are going to remedy this pathology of the NCA by {\it adding} one-body 
self-energy correction to the self-energy $\Sigma_S(\omega)$ to satisfy
the Fermi liquid relation. The self-energy for the dot $S$ can be written 
as a sum of two contributions
\begin{eqnarray}
\Sigma_S^{r} (\omega) 
 &=& \Sigma_{\rm hyb} ^{r} (\omega) + \Sigma_U^{r} (\omega), ~~
\Sigma_{\rm hyb} ^{r} (\omega) ~=~ |W|^2 \tilde{D}^{r} (\omega). 
\end{eqnarray} 
The self-energy $\Sigma_{NCA}^{r} (\omega)$ calculated from the NCA is known to 
show the correct energy and temperature dependence near $\omega=0$ and $T=0$.
(Temperature must be higher than the pathogical temperature estimated in, 
e.g., the references\cite{eqnca1}.) 
We assume that $\Sigma_U(\omega)$ is well-captured in the NCA.
Since $- \mbox{Im} \Sigma_U^{r} (\omega) \propto \omega^2 + [\pi k_BT]^2$ 
for $\omega, T < T_K$ (look at the references\cite{fliquid}), 
we add the correction term to $\Sigma_{NCA}$
in order to make $D_{A}^{r} (\omega)$ obey the causality relation. 
\cite{correction}
\begin{eqnarray}
\Sigma_{S}^{r} (\omega) 
 &\approx& \Sigma_{NCA}^{r} (\omega) + (|W|^2 - W_{NCA}^2) \tilde{D}_{A}^{r} (\omega).
\end{eqnarray} 
Here $W_{NCA}^2$ is determined numerically from the calculated 
$\Sigma_{NCA}^{r} (\omega)$ at the lowest possible temperature. 
\begin{eqnarray}
{W_{NCA}^2 \over \Gamma} 
 &\equiv& - \mbox{Im} \Sigma_{NCA}^{r} (0). 
\end{eqnarray}

\newpage
%
%

\end{document}